# Nitrogen magneto-ionics

Julius de Rojas[1], Alberto Quintana[2], Aitor Lopeandía[1], Joaquín Salguero[3], Beatriz Muñiz[3], Fatima Ibrahim[4], Mairbek Chshiev[4], Maciej O. Liedke[5], Maik Butterling[5], Andreas Wagner[5], Veronica Sireus[1], Llibertat Abad[6], Christopher J. Jensen[2], Kai Liu[2], Josep Nogués[7,8], José L. Costa-Krämer[3], Enric Menéndez[1*] and Jordi Sort[1,8*]

[1]Departament de Física, Universitat Autònoma de Barcelona, E-08193 Cerdanyola del Vallès, Spain

[2]Department of Physics, Georgetown University, Washington, D.C. 20057, United States

[3]IMN-Instituto de Micro y Nanotecnología (CNM-CSIC), Isaac Newton 8, PTM, 28760 Tres Cantos, Madrid, Spain

[4]Univ. Grenoble Alpes, CEA, CNRS, Spintec, 38000 Grenoble, France

[5]Institute of Radiation Physics, Helmholtz-Zentrum Dresden-Rossendorf, Dresden 01328, Germany

[6]Institut de Microelectrònica de Barcelona, IMB-CNM (CSIC), Campus UAB, E-08193 Bellaterra, Spain

[7]Catalan Institute of Nanoscience and Nanotechnology (ICN2), CSIC and BIST, Campus UAB, Bellaterra, E-08193 Barcelona, Spain

[8]Institució Catalana de Recerca i Estudis Avançats (ICREA), Pg. Lluís Companys 23, E-08010 Barcelona, Spain

*Email: enric.menendez@uab.cat (E. Menéndez), jordi.sort@uab.cat (J. Sort)

**Abstract**

So far, magneto-ionics, understood as voltage-driven ion transport in magnetic materials, has largely relied on controlled migration of oxygen ion/vacancy and, to a lesser extent, lithium and hydrogen. Here, we demonstrate efficient, room-temperature, voltage-driven nitrogen transport (*i.e.*, nitrogen magneto-ionics) by electrolyte-gating of a single CoN film (without an ion-reservoir layer). Nitrogen magneto-ionics in CoN is compared to oxygen magneto-ionics in $Co_3O_4$, both layers showing a nanocrystalline face-centered-cubic structure and reversible voltage-driven ON-OFF ferromagnetism. In contrast to oxygen, nitrogen transport occurs uniformly creating a plane-wave-like migration front, without assistance of diffusion channels. Nitrogen magneto-ionics requires lower threshold voltages and exhibits enhanced rates and cyclability. This is due to the lower activation energy for ion diffusion and the lower electronegativity of nitrogen compared to oxygen. These results are appealing for the use of magneto-ionics in nitride semiconductor devices, in applications requiring endurance and moderate speeds of operation, such as brain-inspired computing.

Magneto-ionics[1–11], *i.e.*, the change in the magnetic properties of materials due to electric-field-induced ion motion, is acquiring a leading role, among other magnetoelectric mechanisms (intrinsic[12] or extrinsic[13] multiferroicity, electric charge accumulation[14–16]), to control magnetism with voltage[17,18]. This is triggered by its capability to largely modulate magnetic properties in a permanent and energy-efficient way[1,19]. Usually, magneto-ionic systems comprise layered heterostructures built around a ferromagnetic target material, such as Co or Fe, grown adjacent to solid-state electrolyte films (*e.g.*, $GdO_x$[1,8] or $HfO_2$[20]). Depending on the voltage polarity, these electrolytes accept or donate oxygen, acting as ion reservoirs. In this way, for instance, the effective magnetic easy axis of ferromagnetic layers can be precisely controlled[1,8]. However, room-temperature ionic response is slow (~$10^3$ s[1]). Therefore, being ion migration a thermally activated process[21], ion diffusion is commonly assisted with heat[1,2], which is detrimental in terms of energy efficiency. Sometimes cyclability is limited due to cumulative irreversible changes that the target materials undergo from structural/compositional viewpoints[1,2]. Recently, through a proton-based route, $10^{-3}$ s ionic motion has been demonstrated at room temperature, with good endurance, despite restricted hydrogen retention[7]. Hydrogen is mainly adsorbed rather than absorbed, which imposes stringent limitations on the thickness of the ferromagnet. Moreover, the need for $GdO_x$ and $HfO_x$ as ion reservoirs is to some extent undesirable since both Gd and Hf are considered "critical raw materials" and their use in devices is discouraged.[22] Other approaches relying on the insertion/removal of ions, such as $Li^+$, into a ferromagnet are promising in terms of reversibility. However, operation speeds are often slow and, due to incompatibilities with CMOS architectures, applications in electronics are limited[23–25].

An alternative approach is to use target materials which are already oxidized. Magneto-ionics using structural oxygen (*i.e.*, oxygen incorporated in the crystallographic structure of the actuated material) exhibits outstanding stability and reversibility[6,11]. This has been demonstrated in electrolyte-gated[14,26–30], thick (≥ 100 nm) paramagnetic $Co_3O_4$ films, in which room-temperature voltage-controlled ON-OFF ferromagnetism has been achieved, benefiting from defect-assisted voltage-driven transport of structural oxygen[6]. Nevertheless, there is an inherent voltage trade-off between induced magnetization, speed and cyclability. Specifically, the generated magnetization increases with voltage, whereas cyclability degrades for exceedingly high voltages due to irreversible losses of oxygen from the electrolyte (*i.e.*, bubbling)[6,11].

Here, efficient nitrogen magneto-ionics is demonstrated as an alternative to oxygen magneto-ionics. CoN and $Co_3O_4$ single-layer films are voltage-actuated to compare nitrogen *vs*. oxygen magneto-ionic performances. These materials were selected since they both exhibit voltage-induced ON-OFF ferromagnetic transitions. Both films were grown by sputtering, have the same thickness and exhibit a similar (nanocrystalline, face-centered cubic) microstructure. Remarkably, voltage-driven transport of structural nitrogen is energetically more favorable than oxygen, resulting in lower operating voltages and enhanced cyclability. This together with the lower electronegativity (*i.e.*, weaker bonds with Co) of nitrogen with respect to oxygen leads to overall enhanced magneto-ionic effects. Controlled motion of nitrogen ions with voltage might enable the use of magneto-ionics in new technological areas that require endurance and moderate operation speeds (*e.g.*, neuromorphic computing[31] or micro-electro-mechanical systems[32]).

**Structural, magnetic and transport characterization**

The investigated heterostructures comprise (i) 85 nm $Co_3O_4$ and (ii) 85 nm CoN, grown onto 60 nm Cu/20 nm Ti/0.5 mm [100]-oriented Si. The $\theta/2\theta$ X-ray diffraction (XRD) patterns of the as-prepared $Co_3O_4$ and CoN films (Supplementary Fig. S1a) are consistent with the presence of textured and polycrystalline $Fd\bar{3}m$ $Co_3O_4$ and $F\bar{4}3m$ CoN phases. From the XRD peak positions of CoN, a slightly expanded lattice parameter is obtained ($a$ = 4.403 Å, *i.e.*, 3% larger than the tabulated value), suggesting the presence of interstitial nitrogen. In fact, the relative Co (49.2 %): N (50.8 %) atomic ratio, obtained by electron energy loss spectroscopy, indicates a small off-stoichiometry in the film's composition. Using XRD Rietveld refinement, the crystallite size in both systems is estimated to be about 25 nm. Moreover, high resolution transmission electron microscopy (TEM) was performed in the cross-sections of as-prepared CoN and $Co_3O_4$ films (Supplementary Fig. S1b and S2). The fast Fourier transform (FFT) analysis of the TEM images is consistent with CoN (Supplementary Table S1). Similarly, the FFT analysis in Fig. S2 gives spots whose positions are consistent with $Co_3O_4$ (Supplementary Table S2)[6].

Importantly, the as-prepared $Co_3O_4$ and CoN films exhibit no appreciable ferromagnetic behavior (black loops in Figs. 1b and 1c, Supplementary Fig. S3), in agreement with their room-temperature paramagnetic nature. Moreover, the resistivity values at room temperature of $Co_3O_4$ (19 $\Omega$ cm) and CoN (4.11 × $10^{-4}$ $\Omega$ cm) (Supplementary Fig. S4) reveal their insulating and semiconductor character, respectively. This allows the layers to hold electric fields across them, which is a necessary condition for magneto-ionic phenomena.

**Oxygen *vs.* nitrogen magneto-ionics: magnetoelectric characterization**

The voltage actuation is carried out via electrolyte-gating using an electrochemical capacitor configuration[14,26–30]: the Cu buffer layer and a Pt wire act as working and counter electrodes, respectively (Fig. 1a). This way, the overall film area exposed to the liquid electrolyte[29] is activated, establishing a well-defined out-of-plane electric field[11]. To investigate magneto-ionic motion, the films were electrolyte-gated at –50 V for 12 hours and, during this time, consecutive magnetic hysteresis loops (25 min duration each) were sequentially recorded. In Figs. 1b and 1c, the red loops correspond to the first measurement under voltage and the black arrows indicate the first sweeping leg of the cycle. In this timeframe (during the descending branch of the loop), the magnetization ($M$) of the CoN film significantly increases, whereas $Co_3O_4$ still remains paramagnetic (*i.e.*, with no permanent $M$), evidencing that nitrogen motion is significantly faster than oxygen transport. This is demonstrated in Fig. 1d, which shows the saturation magnetization, $M_S$, as a function of time $t$ (see Supplementary Fig. S5 for details on $M_S$ quantification). $M_S$ evolves monotonically with time for both films, reaching, after magneto-ionic motion has stabilized, maximum values of 588 and 637 emu cm$^{-3}$ for Co$_3$O$_4$ and CoN, respectively (Table 1). By linearly fitting $M_S$ vs. $t$ during the first minutes of voltage application (wherein $M_S$ in CoN fully saturates), magneto-ionic motion rates of 467.1 and 2602.4 emu cm$^{-3}$ h$^{-1}$ are obtained for oxygen in Co$_3$O$_4$ and nitrogen in CoN, respectively (Table 1). For both systems, when gating at –50 V, bubbling (O$_2$ in Co$_3$O$_4$ and N$_2$ in CoN) occurs. This effect, which results in noisy hysteresis loops, is more pronounced for CoN, where the magneto-ionic response is stronger. The hysteresis cycles from electrolyte-gated CoN are more square-shaped than for Co$_3$O$_4$ (Fig. 1b,c). They exhibit larger squareness [defined as the ratio between the remnant magnetization ($M_R$) and $M_S$ ($M_R/M_S$)], and larger slopes at the coercivity ($H_C$) normalized to $M_S$ [d$M$/d$H$ ($H = H_C$) $M_S^{-1}$] than the loops of Co$_3$O$_4$ throughout the time the voltage was applied (Fig. 1e).

Additionally, the $H_C$ corresponding to Co$_3$O$_4$ scales monotonically with time, while in CoN $H_C$ shows a maximum at the first initial stages of gating and decreases afterwards (Supplementary Fig. S6). The case of CoN bears a resemblance to the characteristic dependence of $H_C$ with particle size in magnetic systems, consistent with a homogeneous generation of ferromagnetic regions, uniformly evolving in size, starting from a superparamagnetic behavior (zero $H_C$), followed by a single domain state (maximum $H_C$) and, afterwards, reaching a multi-domain state with reduced $H_C$[33]. The squareness of the loops and the rather small $H_C$ for CoN indicate an in-plane anisotropy and likely a reversal by domain wall motion, which hints at the uniformity of the generated metallic Co phase (Table 1). Such low coercivities might be also associated with a highly nanostructured or even amorphous-like Co[34]. In contrast, the generated ferromagnetism in Co$_3$O$_4$ shows much larger $H_C$, which results from the formation of isolated Co clusters immersed in a residual Co$_3$O$_4$ matrix[6,11].

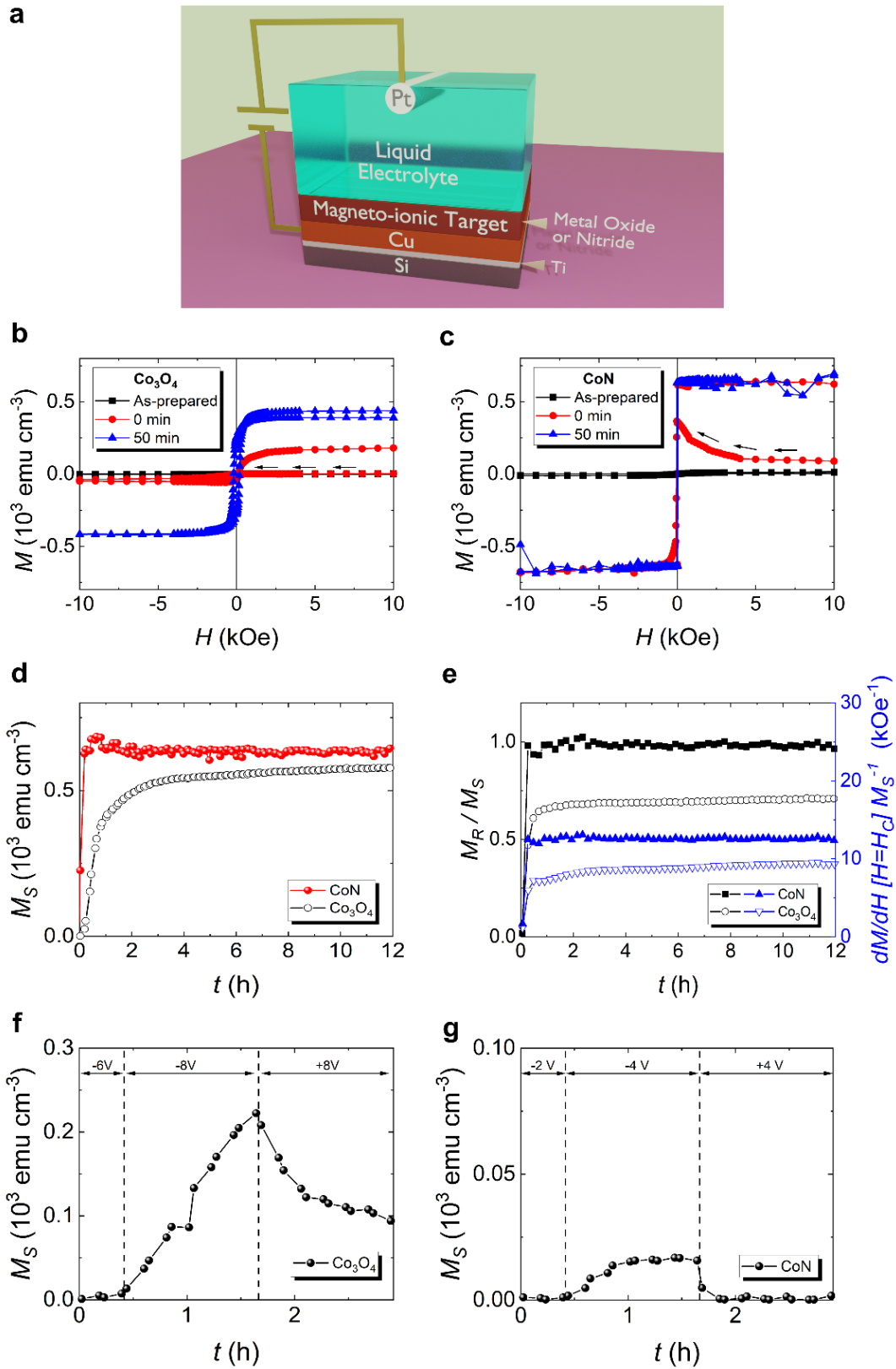

**Fig. 1 | Oxygen *vs*. nitrogen magneto-ionics: magneto-electric characterization. a**, Schematic representation of the electrolyte-gating process in the Co$_3$O$_4$ and CoN films. **b** and **c**, Consecutive hysteresis loops (each of 25 min of duration) under −50 V gating for the Co$_3$O$_4$ and CoN films, respectively, obtained by in-plane vibrating sample magnetometry. **d**, Time evolution of the saturation magnetization ($M_S$ *vs*. $t$) and **e**, squareness ($M_R/M_S$) and slope of hysteresis loop at $H_C$ normalized to $M_S$ (d$M$/d$H$[$H$=$H_C$] $M_S^{-1}$) for each film. **f** and **g**, Time evolution of $M_S$ when the gating is monotonically increased in steps of −2 V to determine the onset voltage required to display ferromagnetism for Co$_3$O$_4$ and CoN, respectively.

The onset voltage for magneto-ionic motion was determined in as-prepared films by monotonically increasing the absolute value of the negative gating in steps of –2 V to observe when the films started to display ferromagnetism (Figs. 1f and 1g, Supplementary Fig. S3). Interestingly, the onset voltage for CoN (–4 V) is clearly lower than for $Co_3O_4$ (–8 V). To investigate reversibility of the magneto-ionic process, both systems were kept at their respective onset voltages for an hour and, then, the polarity was inverted (*i.e.*, +4 and +8 V). Remarkably, while $Co_3O_4$ recovers only partially at +8 V, CoN fully recuperates the pristine paramagnetic state at +4 V. These results anticipate a higher activation energy for oxygen transport than for nitrogen. In fact, to recover the paramagnetic state in $Co_3O_4$ +20 V are required (Table 1). The need to actuate at a higher voltage induces irreversible losses of oxygen since the liquid electrolyte has a limited solubility of oxygen[35]. Note that if –50 V are applied, none of the treated films is recoverable in agreement with the irreversible loss of oxygen and nitrogen through bubbling.

**Table 1 | Oxygen *vs.* nitrogen magneto-ionics by magnetoelectric measurements.** Onset and recovery voltages, magneto-ionic speed and magnetic properties of the generated ferromagnetic phases.

| Film | Onset Voltage (V) | Recovery Voltage (V) | dM/dt @ –50 V (emu cm$^{-3}$ h$^{-1}$) | $M_S$ (emu cm$^{-3}$) | $M_R/M_S$ (%) | dM/dH $M_S^{-1}$ @ $H_C$ (kOe$^{-1}$) | $H_C$ (Oe) |
|---|---|---|---|---|---|---|---|
| $Co_3O_4$ | –8 | +20 | 467.1 | 588 | 62 | 7.4 | 185 |
| CoN | –4 | +4 | 2602.4 | 637 | 96 | 12.2 | 17 |

**Oxygen *vs.* nitrogen magneto-ionics: cyclability**

To investigate the cyclability, the CoN and $Co_3O_4$ films were subjected to –4 V/+4 V and –8 V/+8 V pulses of relatively short duration ($\approx$ 8.5 min/cycle). The duration of the onset negative voltage pulse was selected in each material to give a $\Delta M$ of approximately 1 emu cm$^{-3}$ in the first cycle. This resulted in times of approximately 4.1 min for CoN and 6.2 min for $Co_3O_4$ (evidencing that the response of $Co_3O_4$ is delayed with respect to that of CoN). As seen in Fig. 2, cycling at –/+4 V results in a very stable periodic response for CoN. Conversely, at this voltage, no traces of magneto-ionic effects are observed in $Co_3O_4$, corroborating the need of a higher onset voltage. Good cyclability is observed for $Co_3O_4$ at –8 V/+8 V. Note that, in contrast to Fig. 1f, short pulses at –/+8 V allow recovering the initial state in $Co_3O_4$ after each cycle. Both materials show no progressive irreversible gain/loss of magnetic signal (*i.e.*, no drift) upon successive cycling. However, while the $\Delta M$ amplitude in each cycle remains stable for CoN, a loss of about 25% is observed in $Co_3O_4$ after the first cycle. Overall, Fig. 2 corroborates that magneto-ionic rates, onset voltages and endurance are better when using nitrogen than oxygen migration.

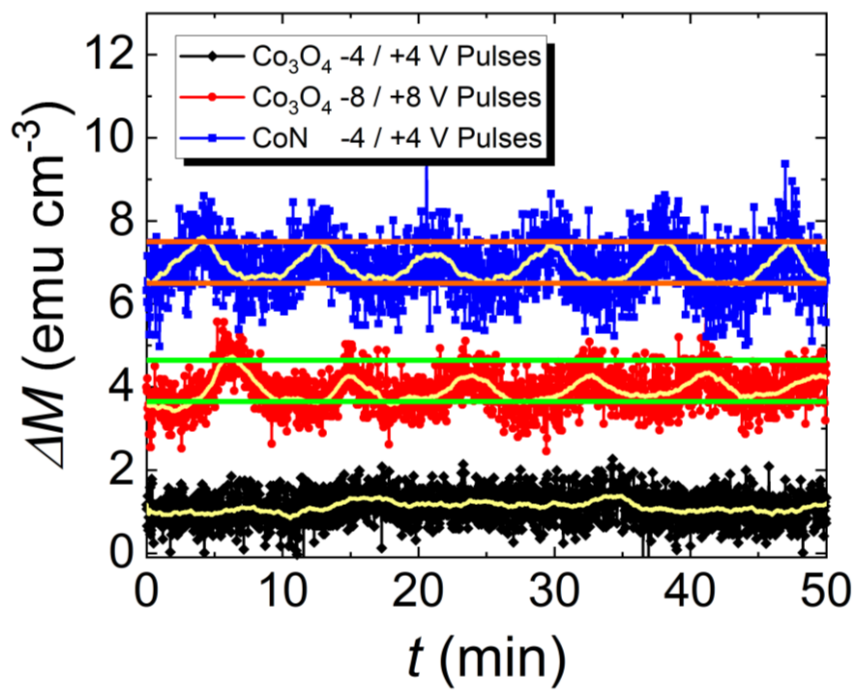

**Fig. 2 | Cyclability**. Magneto-ionic cyclability of the $Co_3O_4$ and CoN films subjected to –4 V/+4 V and –8 V/+8 V, and to –4 V/+4 V, respectively, with pulses of short duration ($\approx$ 8.5 min/cycle). The pale yellow lines represent the average curves. Horizontal lines (red, green) are guides to the eye and span 1 emu cm$^{-3}$.

**Oxygen *vs*. nitrogen magneto-ionics: ion transport mechanisms**

To assess the degree of structural and compositional change that $Co_3O_4$ and CoN undergo upon gating, cross-section lamellae of the as-prepared and electrolyte-gated samples treated at –50 V for 75 min were studied by TEM (Fig. 3). The as-prepared $Co_3O_4$ shows regular, columnar-shaped grains (Fig. 3a, Supplementary Fig. S7a). Co (red) and O (blue) are homogeneously distributed in the as-grown film (Fig. 3b). Conversely, upon gating at –50 V, this morphology drastically changes redistributing the elements into O-rich channels in agreement with previously reported results[6,11] (Figs. 3c and 3d, Supplementary Fig. S7c). This confirms that oxygen transport takes place via a two-fold mechanism: (i) uniform oxygen transport towards the electrolyte and (ii) localized oxygen migration along diffusion channels. Remarkably, high resolution TEM (Supplementary Fig. S8) shows remaining $Co_3O_4$ upon treatment at –50 V (Supplementary Table S3).

The as-prepared CoN film shows an isotropic and highly nanostructured morphology with homogeneous composition (Figs. 3e and 3f, Supplementary Fig. S7b). Remarkably, upon gating at –50 V for 75 min, a well-defined interface (resembling a diffusion front) parallel to the surface is distinguished, dividing the film in two sublayers with different microstructure (Supplementary Fig. S7d and Fig. S9). No traces of nitrogen are detected in any of the two sublayers, evidencing a full denitriding process, consistent with the magnetoelectric characterization. While sublayer 2 is nanocrystalline, sublayer 1 (which was in contact with the electrolyte during magnetoelectric measurements) is amorphous-like and exhibits rather low density (note the decrease of Co signal in Fig. 3h), possibly with a large amount of free volume or nanoporosity. This suggests a complex denitriding process in which, under large negative voltages, not only nitrogen migrates from the sample to the electrolyte, but also Co gets redistributed across the film thickness.

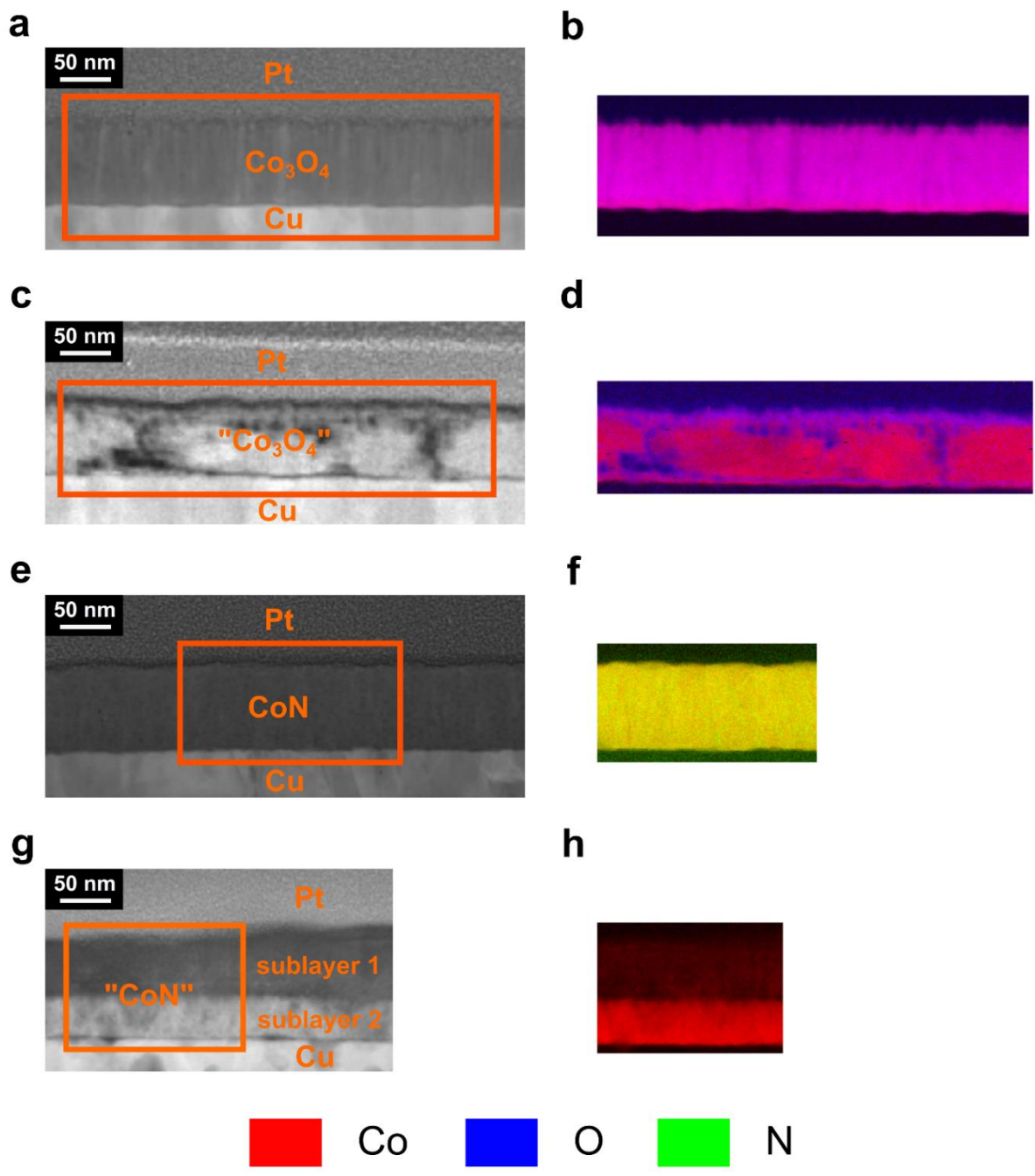

**Fig. 3 | Compositional characterization by high-angle annular dark-field scanning transmission electron microscopy (HAADF-STEM) and electron energy loss spectroscopy (EELS). a-b**, **c-d**, **e-f** and **g-h** are the HAADF-STEM images and corresponding elemental EELS mappings of the areas marked in orange, respectively, of the as-prepared $Co_3O_4$ film, $Co_3O_4$ film subjected to a −50 V for 75 min, as-prepared CoN film and CoN film subjected to a −50 V for 75 min, respectively. The colors corresponding to each element for the EELS analyses are depicted at the bottom of the figure.

To further asses the microstructure of the films upon magneto-ionic actuation, variable energy positron annihilation lifetime spectroscopy (VEPALS) experiments[36-42] were carried out (see

Supplementary Information for details). A plateau at $E_p$ > 5 keV can be observed for all parameters in Fig. 4, which corresponds to the film/buffer (Cu-Ti) interface. For larger $E_p$ the signal from the substrate emerges. In the as-prepared CoN sample, only $\tau_1$ and $\tau_2$ lifetime components are observed, indicating the absence of void-like structures (no $\tau_3$) in the pre-biasing state. $\tau_1$ is around 0.28 ns in the top half of the film, which could correspond to a cluster of more than 4 vacancies[6]. For the bottom half, $\tau_1$ is slightly lower, indicating clusters with less than 4 vacancies. The second lifetime $\tau_2$ represents a mixture of signals from surface states and grain boundaries. In the first top 15 nm, $\tau_2$ is larger than 0.5 ns, indicating the presence of small voids. As seen in Fig. 4b, the relative intensity $I_1$ decreases, while $I_2$ increases reaching similar intensities at the interface with the buffer layer, reflecting the influence of the buffer polycrystallinity in the CoN growth, whose extent decreases with film thickness.

Negative biasing of −20 V and −50 V increases $\tau_1$ and $\tau_2$ and this increase scales with voltage, indicating that the initial open volumes become larger. Already for −20 V, $\tau_3$ emerges, showing the presence of large voids as it happens in $Co_3O_4$[6]. However, for the $Co_3O_4$ system, a monotonic increase of relative intensity $I_2$ across the film thickness is observed[6], whereas relative maxima (marked with arrows in Fig. 4b) are found for CoN. The depth position of these relative maxima increases with the applied voltage, suggesting the occurrence of an interface migration front, in agreement with the TEM observations. This further confirms that, in contrast to oxygen magneto-ionics in $Co_3O_4$, ion transport in CoN takes place via a single and front-like planar wave migration mechanism. For −50 V, the migration front moves deeper into the film in agreement with a more intense denitriding/amorphization process. $I_3$ tends to decrease and vanishes with thickness, evidencing that larger voids are only present at the top surface. The ionic transport upon electrolyte-gating in CoN is thus consistent with a uniform nitrogen migration through vacancies and grain boundaries ($\tau_1$ and $\tau_2$ increase), leaving behind larger grain boundaries and voids.

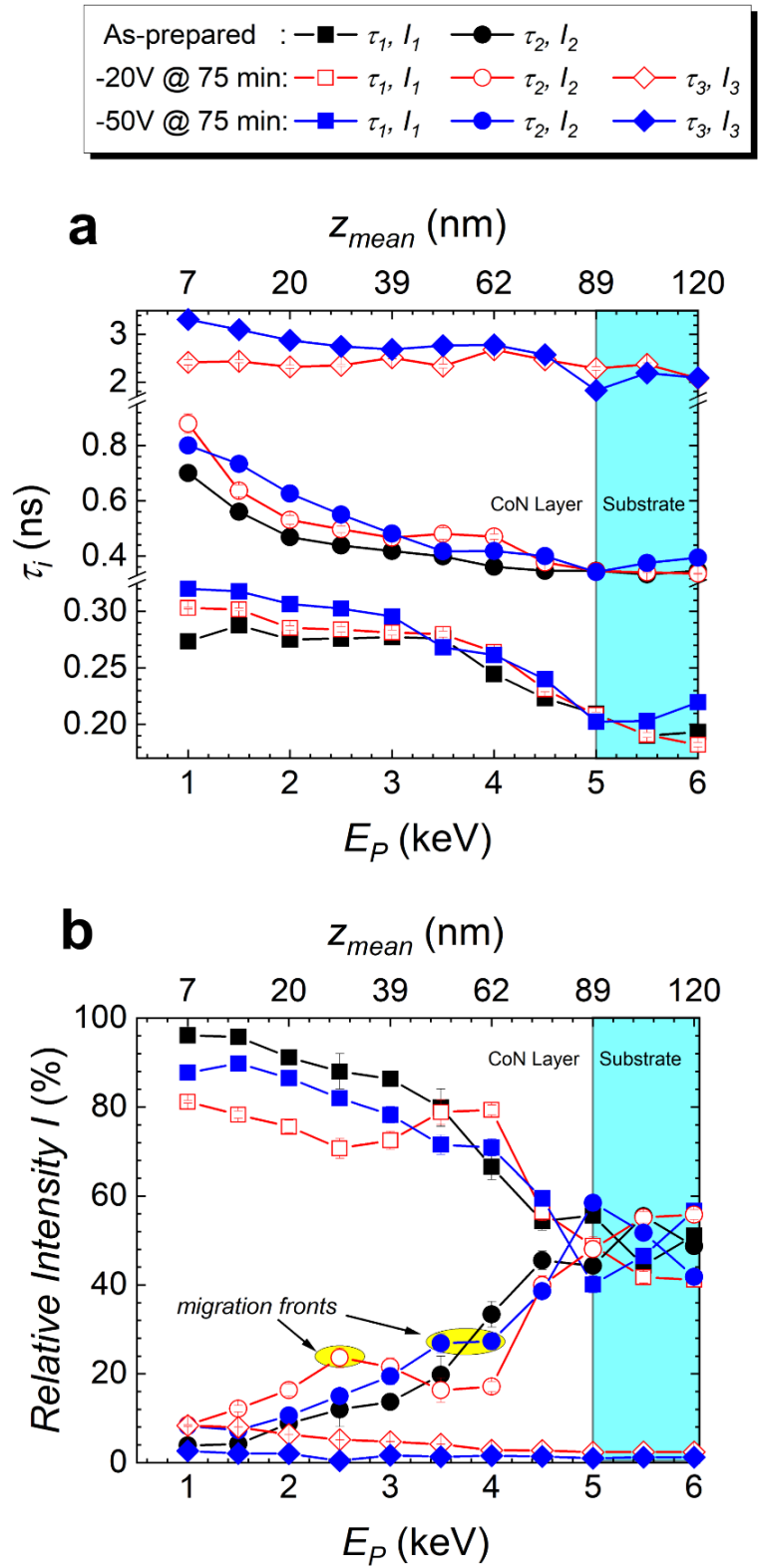

**Fig. 4 | Defect characterization by variable energy positron annihilation lifetime spectroscopy (VEPALS). a** Positron lifetime components $\tau_{i=1-3}$ and **b** their relative intensities $I_{i=1-3}$ as a function of positron implantation energy $E_p$ and mean positron implantation depth $z_{mean}$ for as-prepared and –20 V and –50 V biased CoN films. The non-monotonic change of intensity $I_2$ with depth is linked to the position of the interface between Co sublayers reminiscent of an ionic migration front as highlighted in **b**.

**Oxygen *vs*. nitrogen magneto-ionics: Co-O *vs*. Co-N formation energy**

Neither $Co_3O_4$ nor CoN are ferromagnetic at room temperature. The magnetic properties of these systems are strongly correlated to the amount of either oxygen or nitrogen in the films. According to the virtual crystal approximation, for CoN, beyond 50 at. % of nitrogen in the unit cell, the magnetic moment becomes negligible (ferromagnetism vanishes, Fig. 5a). This explains why the as-prepared CoN film is not magnetic and it also sets a limit for irreversible losses of nitrogen beyond which successive voltage-driven ON-OFF-ON ferromagnetism would be compromised.

To simulate the Co-O and Co-N formation energy (which occurs when the oxygen and nitrogen ions dissolved in propylene carbonate are re-introducing to the films with positive voltage), the insertion of an atom of either oxygen or nitrogen into a cobalt slab has been considered. Minimum energy paths were calculated by the nudged elastic band method (NEB) (Methods). The obtained total energies per atom normalized to the global minimum value are plotted in Fig. 5b as a function of the displacement *z* of the oxygen or nitrogen atom from the cobalt reference layer. The reference *z* = 0 Å is assigned to the position of the outermost cobalt layer. The global energy minimum is found at *z* = 1.14 Å for both oxygen and nitrogen cases. Another local minimum is located around *z* = –0.75 Å for oxygen and *z* = –1 Å for nitrogen. In turn, the calculated energy barriers between the two minima are 1.54 and 1.14 eV/atom for oxygen and nitrogen displacement cases, respectively. Thus, inserting nitrogen into cobalt is energetically more favorable (*i.e.*, less energy required) compared to inserting oxygen. Using the calculated energy barrier values, one can infer the forces acting on the oxygen and nitrogen atoms and, thus, the corresponding critical electric fields, $E_C$, needed to overcome the energy barrier. These electric fields can be estimated as $E_C = \frac{\Delta V}{\Delta z}$ with $\Delta V$ and $\Delta z$ representing, respectively, the electric potential per atom to overcome the energy barrier and the distance between minima that an atom must migrate[19] (for oxygen: $\Delta V$ = 1.54 V and $\Delta z$ = 1.89 Å, and for nitrogen: $\Delta V$ = 1.14 V and $\Delta z$ = 2.14 Å). $E_C$ is found to be 8.1 V/nm and 5.3 V/nm for oxygen and nitrogen migration, respectively, in good agreement with the onset voltages determined from magnetoelectric measurements (Figs. 1f and 1g) bearing in mind the thickness of the electric double layer (1 nm or less).

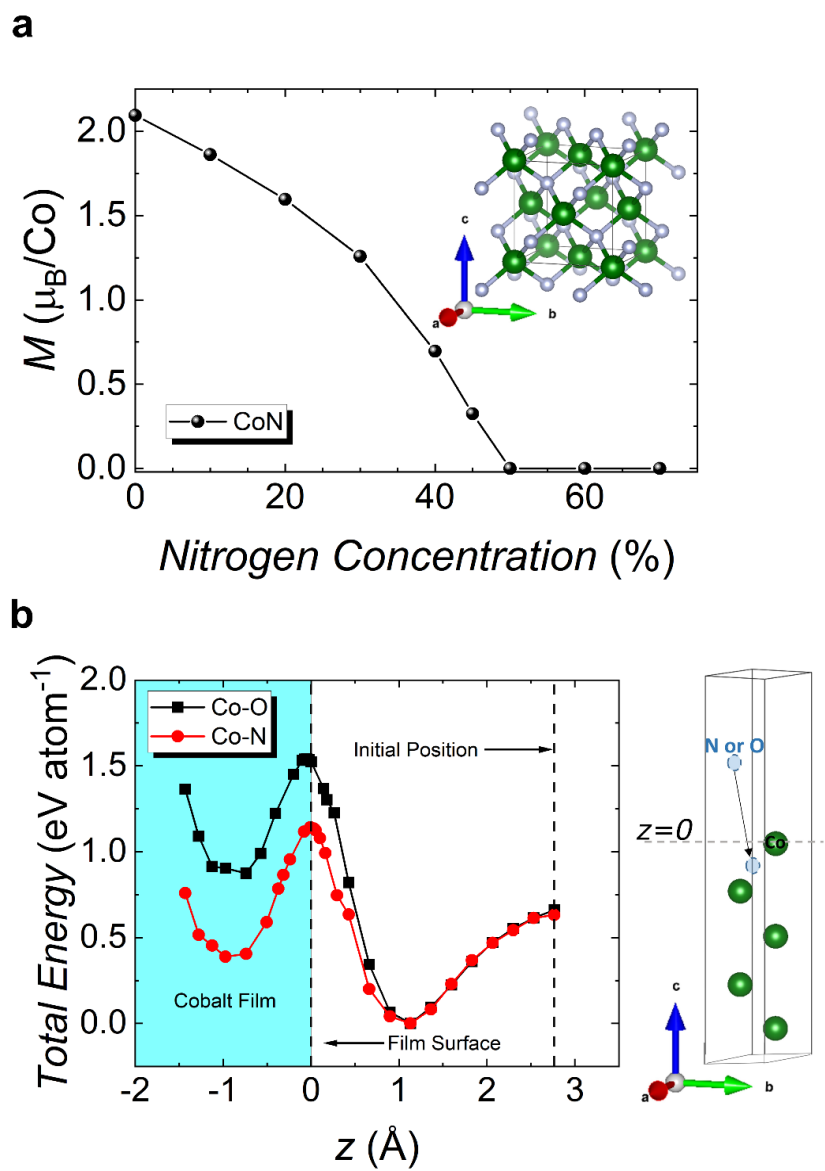

**Fig. 5 | *Ab initio* calculations: magnetism in the Co-N system and Co-O vs. Co-N formation energy. a**, Variation of magnetic moment in Bohr magneton, $\mu_B$, per Co atom as a function of increasing the N percentage in the CoN unit cell, shown in the inset, calculated within Virtual Crystal Approximation (VCA). **b**, Calculated total energy per atom, normalized to the minimum energy value, as a function of the displacement between the reference Co outermost surface atom and the inserted O or N atom. The black squares (red circles) correspond to the O (N) energetic path, respectively. The five-monolayer-thick Co slab is shown in the right panel, where the dashed line indicates the reference $z$ position, which is the outermost Co surface monolayer.

The ionic radii of nitrogen ions are larger than the ionic radius of $O^{2-}$ and, once the energy barrier for ion diffusion is overcome, ionic motion would be, in first approximation, expected to be larger for oxygen than for nitrogen. However, the simulation results indicate the opposite, revealing that other parameters, such as electronegativity, might play a more dominant role than ion size in ionic motion. In fact, the Pauling electronegativity of nitrogen is lower than that of oxygen, resulting in weaker bonds with cobalt and allowing for an enhanced ionic motion.

Our work demonstrates robust room-temperature nitrogen magneto-ionics in semiconducting CoN. Nitrogen magneto-ionics shows reduced activation energies for ionic transport, thus requiring lower voltage actuation. Moreover, the magneto-ionic rates are faster than for oxygen magneto-ionics. This is linked to the conjugation of a lower critical electric field to overcome the energy barrier for ion diffusion and a lower electronegativity of nitrogen with respect to oxygen. Nitrogen magneto-ionics broadens the choice of materials in which voltage-driven effects can be induced and may enable the use of magneto-ionics in devices that require endurance and moderate speeds of operation, such as brain-inspired/stochastic computing or magnetic micro-electro-mechanical systems. The reported effects are particularly appealing to extend the use of nitride semiconductors in new magnetic and spintronic applications.

**Methods**

**Sample preparation**

85 nm thick $Co_3O_4$ and CoN films were grown by reactive sputtering on B-doped, highly conducting [100]-oriented Si wafers (0.5 mm thick), previously coated with 20 nm of Ti and 60 nm of Cu. Depositions were carried always out while partly masking the Cu layer to serve as working electrode.

The $Co_3O_4$ films were grown at room temperature in an AJA International ATC 2400 Sputtering System with a base pressure in the $10^{-8}$ Torr range. High purity Ar and $O_2$ gases were used. The target to substrate distance was around 8 cm and the sputtering rate of about 5 Å/s. $Co_3O_4$ was grown in a 7% $O_2$ / 93% Ar atmosphere at a total pressure of $2.5\times10^{-3}$ Torr.

The CoN films were grown in a home-made triode sputtering system with a base pressure in the $10^{-8}$ Torr range. Ultra-high vacuum was ensured to minimize oxygen contamination and, thus, to rule out traces of oxygen magneto-ionics. The target to substrate distance was around 10 cm and the sputtering rate about 1 Å/s. CoN was grown in a 50% $N_2$ / 50% Ar atmosphere at a total pressure of $8\times10^{-3}$ Torr.

**Magnetoelectric characterization**

Magnetoelectric measurements were carried out by performing vibrating sample magnetometry while electrolyte gating the films in a capacitor configuration at room temperature. A magnetometer from Micro Sense (LOT-Quantum Design), with a maximum applied magnetic field of 2 T, was used. The samples are mounted in a homemade electrolytic cell filled with anhydrous propylene carbonate with $Na^+$ solvated species (5 - 25 ppm). The magnetic properties were measured along the film plane upon applying different voltages, using an external Agilent B2902A power supply, between the Cu working electrode and the counter electrode (Pt wire) in a similar fashion to that presented in references [6], [11], [14] and [28]. The $Na^+$ solvated species in the electrolyte are aimed at reacting with any traces of water[29]. The magnetic signal is normalized to the volume of the sample exposed to the electrolyte. Note that the hysteresis loops were background-corrected using the signal at high fields (*i.e.*, fields always far above saturation fields) to eliminate linear contributions.

**Structural and compositional measurements**

$\theta/2\theta$ X-ray diffraction (XRD) patterns were recorded on a Materials Research Diffractometer (MRD) from Malvern PANalytical company, equipped with a $PIXcel^{1D}$ detector, using Cu $K_\alpha$ radiation. The patterns were analyzed using a full-pattern Rietveld refinement method.

High resolution transmission electron microscopy (HRTEM), high-angle annular dark-field scanning transmission electron microscopy (HAADF-STEM) and electron energy loss spectroscopy (EELS) were performed on a TECNAI F20 HRTEM /STEM microscope operated at 200 kV. Cross sectional lamellae were prepared by focused ion beam and placed onto a Cu transmission electron microscopy grid.

**Transport measurements**

To determine electric properties, both films ($Co_3O_4$ and CoN) were deposited onto high resistivity Si substrates. To assess the semiconducting behavior of CoN, resistivity values were acquired from 30 to 300 K. In all cases, the van der Pauw configuration was used.

**Variable energy positron annihilation lifetime spectroscopy**

Variable energy positron annihilation lifetime spectroscopy (VEPALS) measurements were conducted at the mono-energetic positron source (MePS) beamline, which is an end station of the radiation source ELBE (Electron Linac for beams with high Brilliance and low Emittance) at Helmholtz-Zentrum Dresden-Rossendorf (Germany)[36] using a digital lifetime $CeBr_3$ scintillator detector with a homemade software employing a SPDevices ADQ14DC-2X with 14 bit vertical resolution and 2GS s$^{-1}$ (GigaSamples per second) horizontal resolution and with a time resolution function down to about 0.205 ns. The resolution function required for spectrum analysis uses two Gaussian functions with distinct intensities depending on the positron implantation energy, $E_p$, and appropriate relative shifts. All spectra contained at least 1×10$^7$ counts.

***Ab initio* calculations**

The first-principles calculations were based on the projector-augmented wave (PAW) method[43] as implemented in the VASP package[44–46] using the generalized gradient approximation[47]. We used cubic unit cell with a F$\bar{4}$3m space group for CoN. The full structural relaxation was performed until the forces became smaller than 1 meV Å$^{-1}$, yielding a lattice constant of 4.41 Å. The virtual crystal approximation[48] was used to model the variation of nitrogen per unit cell. To compare the Co-O and Co-N formation energy, the nudged elastic band method (NEB)[49,50] on the oxygen and nitrogen pathway into a five-monolayer thick (0001) hexagonal close-packed Co slab. At each step, the atomic coordinates were relaxed until the forces became smaller than 1 meV Å$^{-1}$. A kinetic energy cut-off of 500 eV was used for the plane-wave basis set and 25×25×25 and 25×25×1 k-point meshes were used to construct the first Brillouin zone in CoN unit cell and the Co slab in the NEB calculations, respectively.

**Data availability**

The data used in this article are available from the corresponding authors upon request.

**Acknowledgements**

Financial support by the European Research Council (SPIN-PORICS 2014-Consolidator Grant, Agreement Nº 648454 and MAGIC-SWITCH 2019-Proof of Concept Grant, Agreement Nº 875018), the Spanish Government (MAT2017-86357-C3-1-R), the Generalitat de Catalunya (2017-SGR-292 and 2018-LLAV-00032), the French ANR (Project "FEOrgSpin") and the European Regional Development Fund (MAT2017-86357-C3-1-R and 2018-LLAV-00032) is acknowledged. ICN2 is funded by the CERCA programme/Generalitat de Catalunya. The ICN2 is supported by the Severo Ochoa Centres of Excellence programme, funded by the Spanish Research Agency (AEI, grant no. SEV-2017-0706). Work at GU has been supported in part by SMART (2018-NE-2861), one of seven centers of nCORE, a Semiconductor Research Corporation program, sponsored by National Institute of Standards and Technology (NIST), and the NSF (DMR-1905468, DMR-1828420). The PALS measurements were carried out at ELBE at the Helmholtz-Zentrum Dresden-Rossendorf e. V., a member of the Helmholtz Association. The authors would like to thank Ahmed G. Attallah and Eric Hirschmann for assistance during the PALS measurements.

**Author contributions**

**E.M.** and **J.S.** had the original idea and led the investigation. **J.d.R.**, **E.M.** and **J.S.** designed the experiments. **A.Q.**, **C.J.J.** and **K.L.** synthesized the $Co_3O_4$ films and **J.S.**, **B.M.**, **J.N.** and **J.L.C.** prepared the CoN films. **E.M.**, **A.Q.** and **K.L.** designed the sample holder to carry out the magnetoelectric measurements. **J.d.R.** and **E.M.** carried out the magnetoelectric measurements and analyzed the data. **J.d.R.**, **A.Q.**, **C.J.J.** and **E.M.** performed the XRD characterization and analyzed the data. **A.L.**, **L.A.** and **E.M.** carried out the TEM and STEM characterization and analyzed the corresponding data. **F.I.** and **M.C**. performed the *ab initio* calculations. **J.S.**, **B.M.**, **J.N.** and **J.L.C.** carried out the transport measurements. **M.O.L.**, **M.B.** and **A.W.** characterized the samples by PALS and analyzed the data. All authors discussed the results and commented on the article. The article was written by **J.d.R.**, **E.M.** and **J.S.**

**Additional information**

Correspondence and requests for materials should be addressed to either **E.M.** or **J.S.**

**Competing financial interests**

The authors declare no competing financial interests.

# SUPPLEMENTARY INFORMATION

## Additional supplementary figures and tables

**Supplementary Fig. S1**

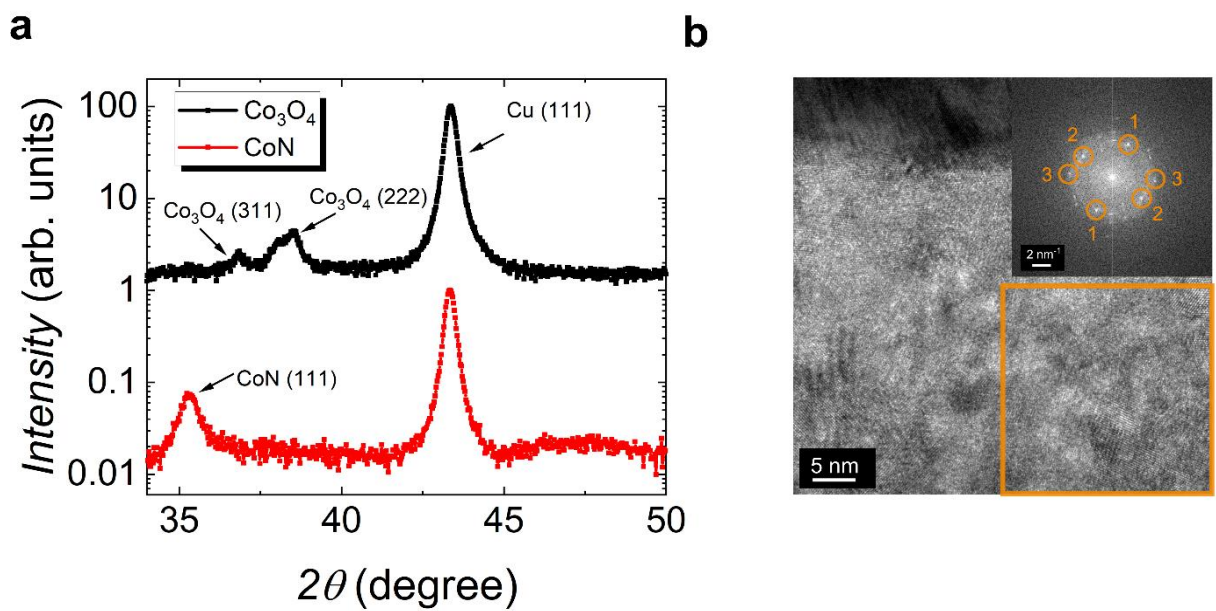

**Fig. S1 | Structural characterization by X-ray diffraction (XRD) and transmission electron microscopy (TEM). a,** $\theta/2\theta$ XRD diffraction patterns of the as-prepared $Co_3O_4$ and CoN films. **b,** High resolution TEM image of the cross-section of an as-prepared CoN film. The inset shows the fast Fourier transform of the area marked with an orange rectangle. For phase identification, the cards no. PDF 00-009-0418 and PDF 00-016-0116, were taken for $Co_3O_4$ and CoN, respectively.

**Supplementary Table S1**

**Table S1 |** Interplanar distances obtained from the spots of the fast Fourier transform in the inset of Fig. S1b. The corresponding Miller indices, according to the PDF 00-016-0116 CoN phase, are also indicated.

| Spot label | Interplanar distance d (Å) | (h k l) planes |
|---|---|---|
| 1 (top right) | 2.56 | (1 1 1) |
| 1 (bottom left) | 2.56 | (1 1 1) |
| 2 (bottom right) | 2.59 | (1 1 1) |
| 2 (top left) | 2.60 | (1 1 1) |
| 3 (right) | 2.20 | (2 0 0) |
| 3 (left) | 2.21 | (2 0 0) |

**Supplementary Fig. S2**

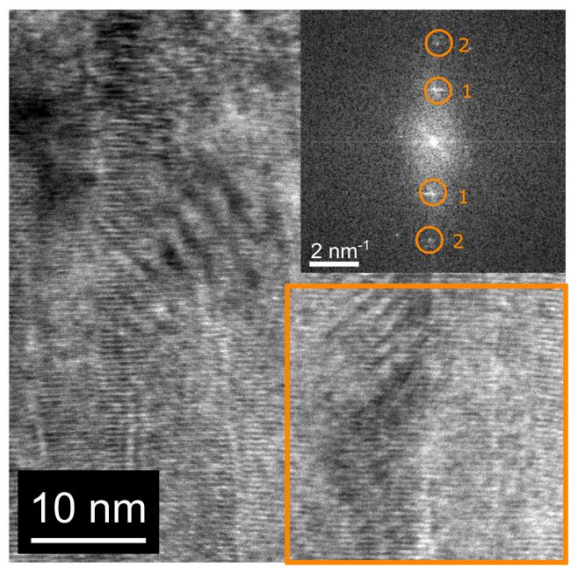

**Fig. S2 | Structural characterization by transmission electron microscopy (TEM).** High resolution TEM image of the cross-section of an as-prepared Co$_3$O$_4$ film. The upper inset show the fast Fourier transform of the area marked with a rectangle.

**Supplementary Table S2**

**Table S2 |** Interplanar distances obtained from the spots of the fast Fourier transform in the inset of Fig. S2a. The corresponding Miller indices, according to the PDF 00-009-0418 Co$_3$O$_4$ phase, are also indicated.

| Spot label | Interplanar distance d (Å) | (h k l) planes |
|---|---|---|
| 1 (middle top) | 4.74 | (1 1 1) |
| 1 (middle bottom) | 4.68 | (1 1 1) |
| 2 (top) | 2.52 | (3 1 1) |
| 2 (bottom) | 2.51 | (3 1 1) |

**Supplementary Fig. S3**

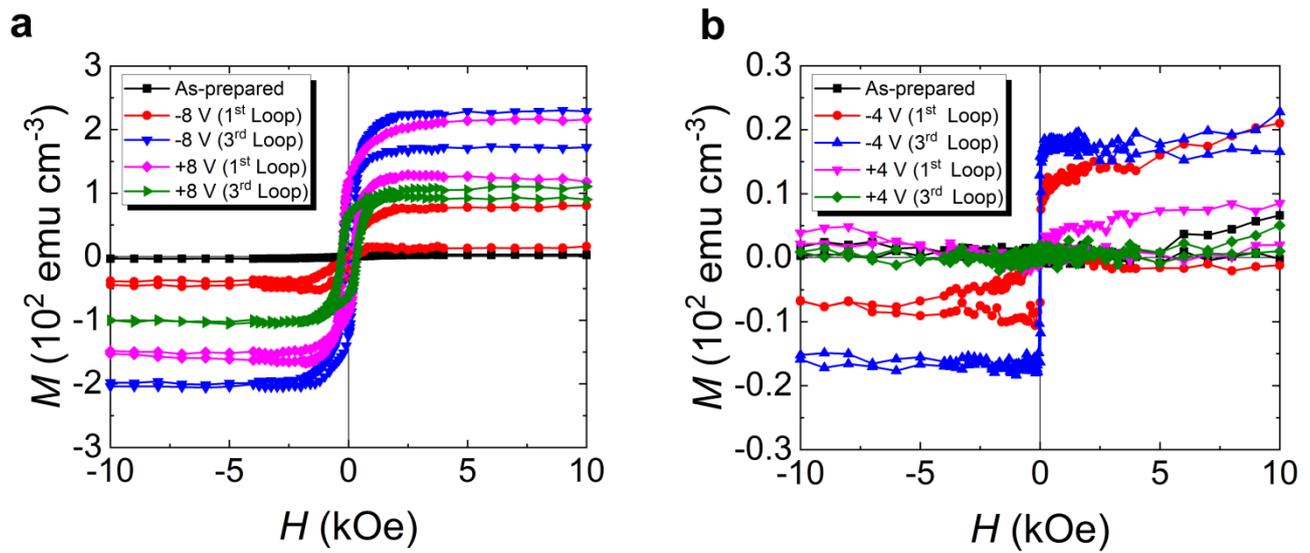

**Fig. S3 | Onset voltage and recovery process**. **a** and **b** consecutive hysteresis loops (each of 25 min of duration) under –8 V and –4 V gating and the corresponding recovery loops recorded at +8 V and +4 V for the $Co_3O_4$- and CoN-based heterostructures, respectively.

**Supplementary Fig. S4**

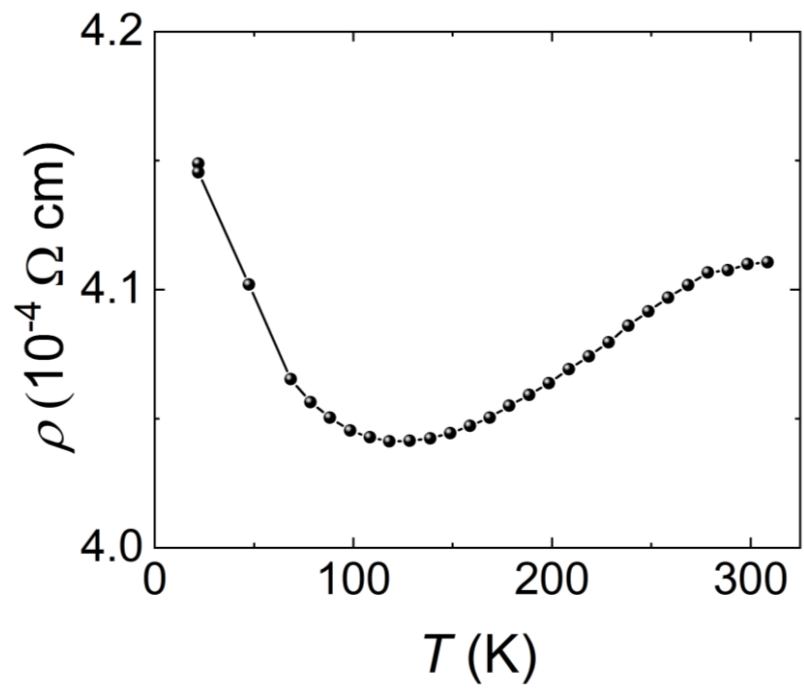

**Fig. S4 | Electric transport properties.** Resistivity as a function of the temperature for the CoN film grown on an insulating substrate.

**Supplementary Fig. S5**

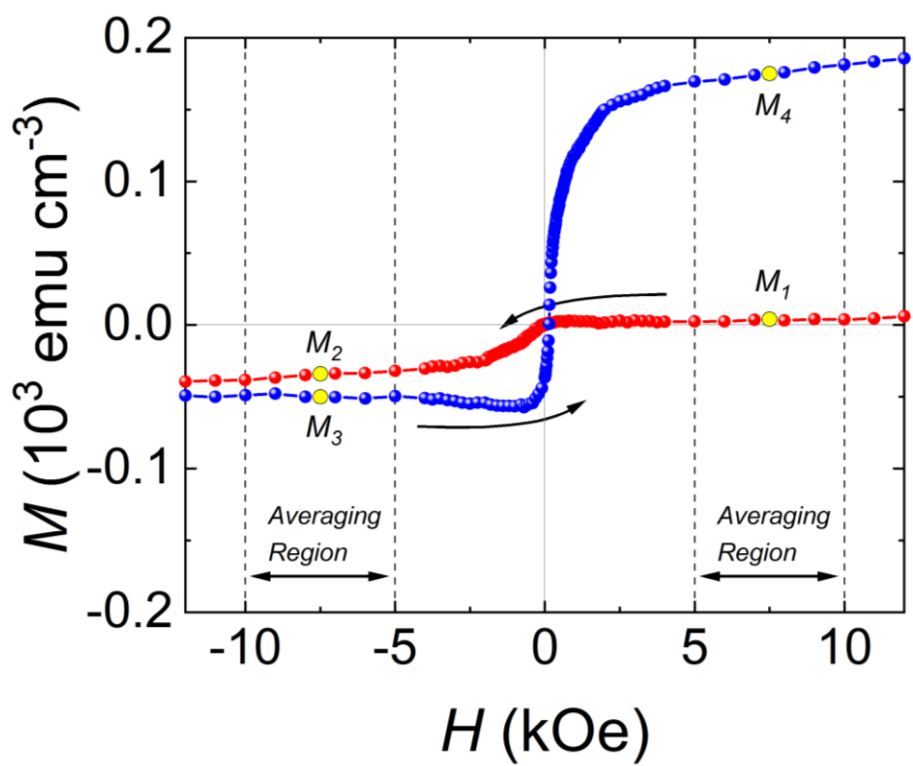

**Fig. S5 | Saturation magnetization ($M_S$) quantification from the hysteresis loops**. The schematic representation is based on the first hysteresis loop of the $Co_3O_4$ film upon electrolyte-gating at −50 V. All loops are previously slope-corrected for any linear contributions. For both descending and ascending branches of the loop (represented by arrows), $M_S$ is calculated in the negative and positive field regions above the anisotropy field: between 5 and 10 kOe in the positive field range, and −10 and −5 kOe in the negative field range. An average of the points is found as the effective $M_S$ (indicated with yellow dots): $M_1$, $M_2$, $M_3$ and $M_4$, and linked to a time during the measurement of the loop. The time corresponding to this averaging region is half a minute.

**Supplementary Fig. S6**

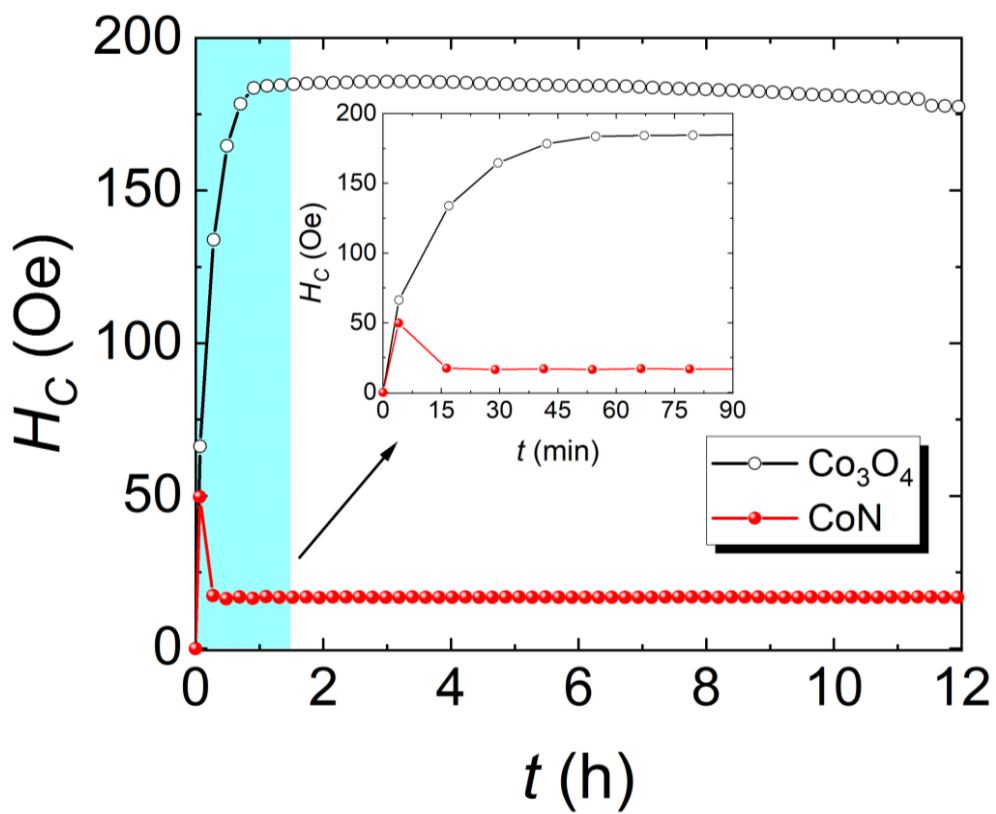

**Fig. S6 | Time evolution of the coercivity during electrolyte-gating.** Evolution of the coercivity ($H_C$) with time for the $Co_3O_4$ and CoN films subjected to a voltage of −50 V for 12 hours. The inset represents zoom of the shaded region (first 90 min).

**Supplementary Fig. S7**

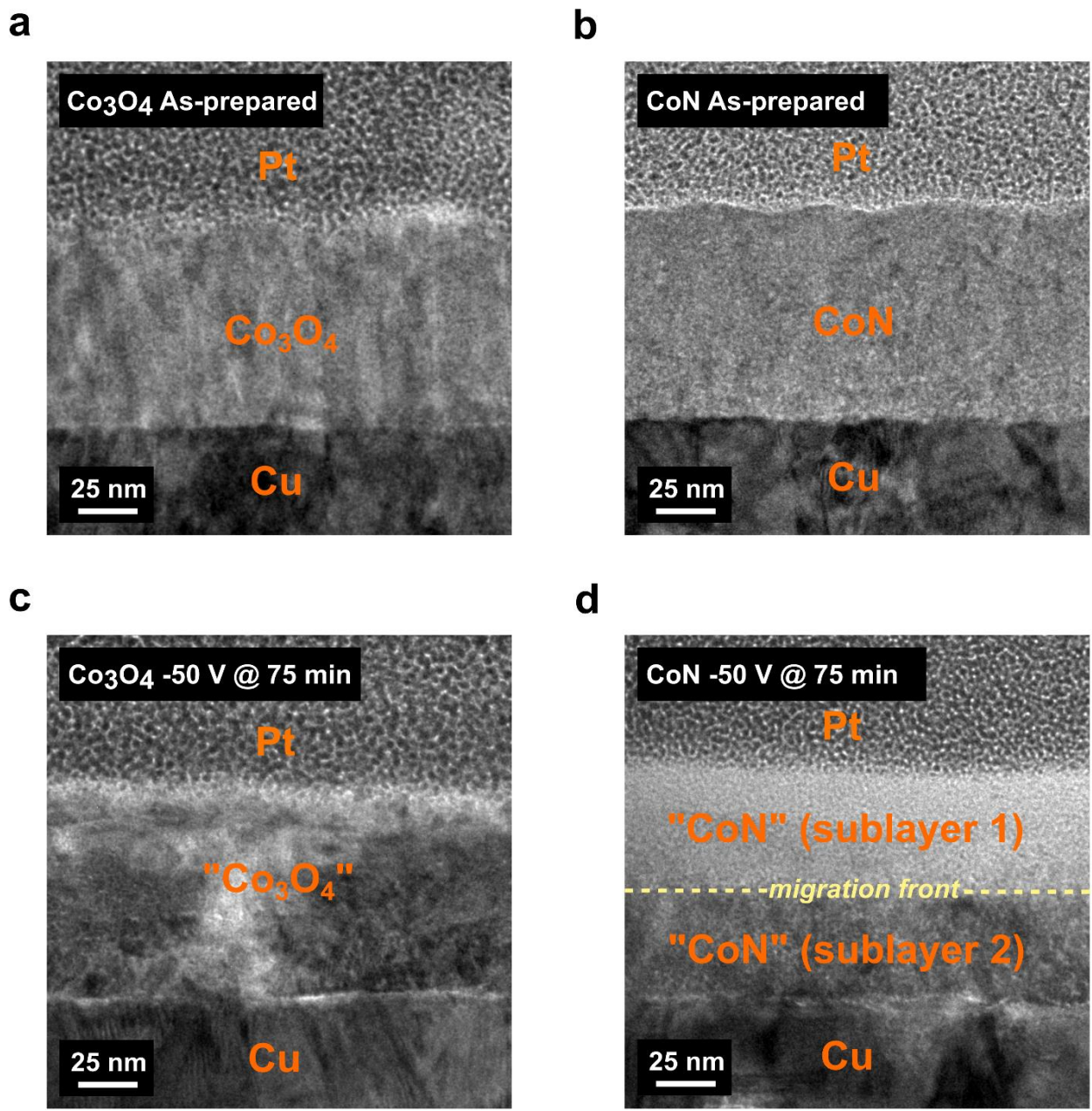

**Fig. S7 | Structural characterization by transmission electron microscopy (TEM). a** and **b** are TEM images of the as-prepared $Co_3O_4$ and CoN films, respectively. **c** and **d** are TEM images of the $Co_3O_4$ and CoN films subjected to –50 V for 75 min, respectively.

**Supplementary Fig. S8**

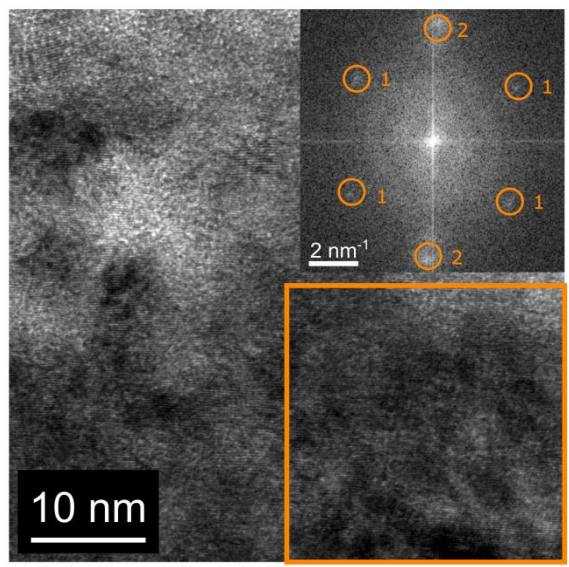

**Fig. S8 | Structural characterization by high resolution transmission electron microscopy (TEM) of the $Co_3O_4$ film after voltage actuation.** High resolution TEM image of the cross-section of a $Co_3O_4$ film treated −50 V for 75 min. The upper inset shows the fast Fourier transforms of the area marked with a rectangle.

**Supplementary Table S3**

**Table S3 |** Interplanar distances obtained from the spots of the fast Fourier transform in the inset of Fig. S8. The corresponding Miller indices, according to the PDF 00-009-0418 $Co_3O_4$, PDF 00-005-0727 hexagonal-closed packed (HCP) Co and PDF 00-015-0806 face-centered cubic (FCC) Co phases, are indicated.

| Spot label | Interplanar distance d (Å) | (h k l) planes |
|---|---|---|
| 1 (top right) | 2.51 | $Co_3O_4$ (3 1 1) |
| 1 (bottom right) | 2.48 | $Co_3O_4$ (3 1 1) |
| 1 (bottom left) | 2.52 | $Co_3O_4$ (3 1 1) |
| 1 (top left) | 2.51 | $Co_3O_4$ (3 1 1) |
| 2 (top) | 2.07 | $Co_3O_4$ (4 0 0)<br>HCP-Co (0 0 2)<br>FCC-Co (1 1 1) |
| 2 (bottom) | 2.06 | $Co_3O_4$ (4 0 0)<br>HCP-Co (0 0 2)<br>FCC-Co (1 1 1) |

**Supplementary Fig. S9**

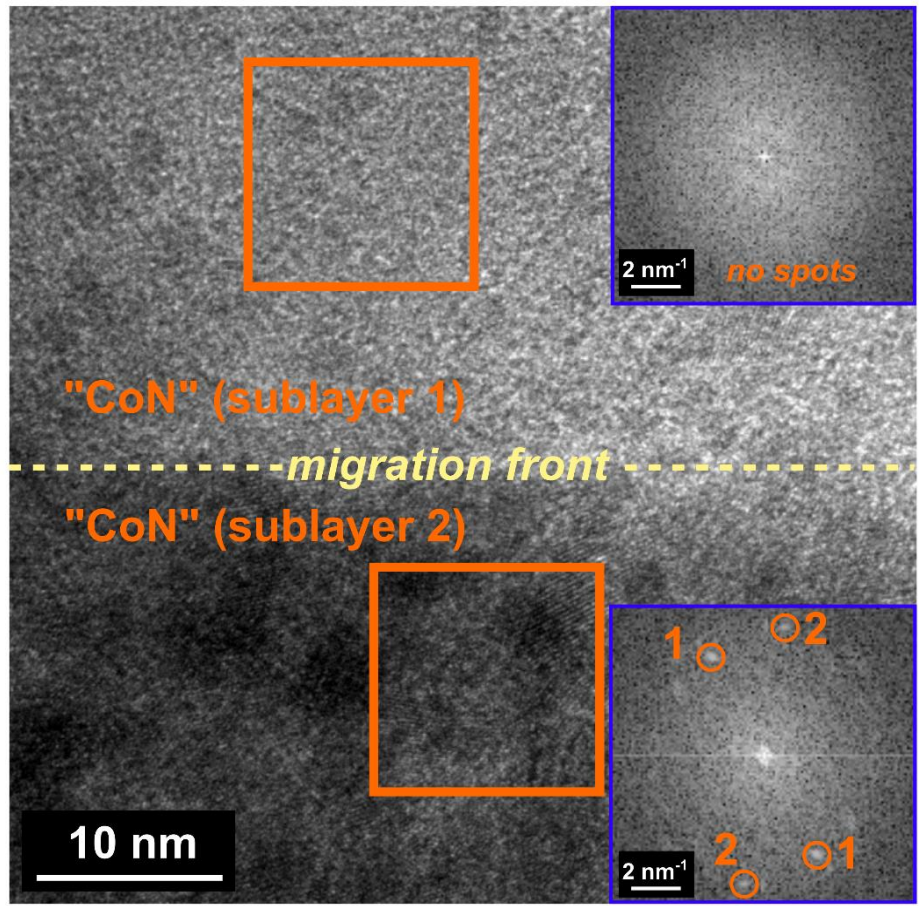

**Fig. S9 | Structural characterization by high resolution transmission electron microscopy (TEM) of the CoN film after voltage actuation.** High resolution TEM image of the cross-section of a CoN film treated −50 V for 75 min. While clear diffraction spots can be seen in sublayer 2 (in contact with Cu), sublayer 1 (in contact with the electrolyte) is amorphous-like. Note that spots "2" likely correspond to HCP-Co (1 0 0), whereas spots "1" match the position of CoO (1 1 1) planes. This oxide probably forms due to natural oxidation of the lamella (Co passivation).

**Additional information on the variable energy positron annihilation lifetime spectroscopy experiments**

Variable energy positron annihilation lifetime spectroscopy (VEPALS) measurements were conducted at the Mono-energetic Positron Source (MePS) beamline at Helmholtz-Zentrum Dresden-Rossendorf, Germany (see Methods for setup technical details)[1]. The spectra were deconvoluted using the non-linearly least-squared based package PALSfit fitting software[2] into discrete lifetime components, which directly confirm different defect types —*i.e.*, sizes— (Fig. 4).

The corresponding relative intensities ($I_i$) reflect to a large extent the concentration of each defect type. In general, positron lifetime ($\tau_i$) is directly proportional to defect size, *i.e.*, the larger the open volume, the lower the probability and the longer it takes for positrons to be annihilated with electrons[3–5]. The positron lifetime and its intensity are probed as a function of positron implantation energy $E_p$ or, in other words, implantation depth (thickness). A mean positron implantation depth $z_{mean}$ can be roughly approximated by a material density dependent formula[6]: $z_{mean}$ (nm) = $(36/\rho(\text{g cm}^{-3}))\cdot(E_p(\text{keV}))^{1.62}$, where $\rho$ = 5.47 g cm$^{-3}$ for CoN was used. The shortest lifetime component ($\tau_1$ < 0.32 ns) represents positron annihilation inside vacancy clusters (likely within grains). The intermediate lifetime (0.35 < $\tau_2$ < 0.90 ns) accounts for annihilation at larger vacancy clusters (linked to grain boundaries), surface states, and small voids/pores (0.28 - 0.37 nm in diameter, calculated based on the shape-free model for pore-size estimation of Wada *et al.*[7]). The longest lifetime component (2.3 < $\tau_3$ < 3.3 ns) indicates contributions of larger voids (0.58 - 0.74 nm in diameter).